\documentclass[11pt,a4paper]{article} 
 \usepackage[margin=1in,footskip=0.25in]{geometry}
 \usepackage[numbers,sort&compress]{natbib}
 \usepackage{amsmath}
\usepackage{amssymb}
\usepackage{authblk}
\usepackage{graphicx}
\usepackage{comment}
\usepackage{braket}
\usepackage{caption}
\usepackage{multirow}
\usepackage[toc,page]{appendix}
\usepackage{amsfonts}
\usepackage{rotating}

\title{Magnetic domains and domain wall pinning in two-dimensional ferromagnets revealed by nanoscale imaging}

\author[1,8]{Qi-Chao Sun}
\author[2,8]{Tiancheng Song}
\author[2]{Eric Anderson}
\author[1]{Tetyana Shalomayeva}
\author[3]{Johaness F\"orster}
\author[1]{Andreas Brunner}
\author[4]{Takashi Taniguchi}
\author[4]{Kenji Watanabe}
\author[3]{Joachim Gr\"afe}
\author[1,5,*]{Rainer St\"ohr}
\author[2,6]{Xiaodong Xu}
\author[1,7]{J\"org Wrachtrup}

\affil[1]{3. Physikalisches Institut, University of Stuttgart, 70569 Stuttgart, Germany}
\affil[2]{Department of Physics, University of Washington, Seattle, Washington 98195, USA}
\affil[3]{Max Planck Institute for Intelligent Systems, 70569 Stuttgart, Germany}
\affil[4]{National Institute for Materials Science, Tsukuba, Ibaraki 305-0044, Japan}
\affil[5]{Center for Applied Quantum Technology, University of Stuttgart, 70569 Stuttgart, Germany}
\affil[6]{ Department of Materials Science and Engineering, University of Washington, Seattle, Washington 98195, USA}
\affil[7]{ Max Planck Institute for Solid State Research, 70569 Stuttgart, Germany}
\affil[8]{These authors contributed equally}
\affil[*]{ e-mail: q.sun@pi3.uni-stuttgart.de; rainer.stoehr@pi3.uni-stuttgart.de}

\begin{document}

\maketitle

Magnetic-domain structure and dynamics play an important role in understanding and controlling the magnetic properties of two-dimensional magnets, which are of interest to both fundamental studies and applications\cite{Huang.2017,Gong.2017,Burch.2018,Mak.2019b,Hubert.1998}. However, the probe methods based on the spin-dependent optical permeability\cite{Huang.2017,Gong.2017,Jiang.2018b} and electrical conductivity\cite{Fei.2018,Deng.2018,Klein.2018,Song.5262019} can neither provide quantitative information of the magnetization nor achieve nanoscale spatial resolution. These capabilities are essential to image and understand the rich properties of magnetic domains. Here, we employ cryogenic scanning magnetometry using a single-electron spin of a nitrogen-vacancy center in a diamond probe to unambiguously prove the existence of magnetic domains and study their dynamics in atomically thin CrBr$_3$. The high spatial resolution of this technique enables imaging of magnetic domains  and allows to resolve domain walls pinned by defects. By controlling the magnetic domain evolution as a function of magnetic field, we find that the pinning effect is a dominant coercivity mechanism with a saturation magnetization of about 26~$\mu_B$/nm$^2$ for bilayer CrBr$_3$. The magnetic-domain structure and pinning-effect dominated domain reversal process are verified by micromagnetic simulation. Our work highlights scanning nitrogen-vacancy center magnetometry as a quantitative probe to explore two-dimensional magnetism at the nanoscale.

\

Two-dimensional (2D) magnetism is of fundamental interest for the study of long-range magnetic order in the presence of the enhanced spin fluctuation at reduced dimensionality\cite{N.D.MerminandH.Wagner.}. Recent developments in van der Waals magnetic materials have greatly enriched the variety and controllability of 2D magnets, and have magnetic properties that can be modified via electron doping\cite{Jiang.2018b,Huang.2018} and by changing layer number\cite{Huang.2017,Gong.2017} or stacking order\cite{Song.5262019,Chen.2019}.  Due to their micrometer size and atomic thickness, the magnetic signal of 2D magnets is too weak to be detected by conventional magnetometry. Several probe techniques, such as magneto-optical Kerr effect microscopy\cite{Huang.2017,Gong.2017}, magnetic circular dichroism microscopy\cite{Jiang.2018b,jinchenhao.}, anomalous hall effect\cite{Fei.2018,Deng.2018}, and electron tunneling\cite{Klein.2018,Song.5262019} have been used in previous studies. Although these methods can reveal phase transitions, quantitative magnetization information is difficult to extract from spin-related signals. Moreover, only micrometer-scale spatial resolution can be achieved by these methods due to the laser diffraction limit or the size of the electrode. Quantitatively study of 2D magnets at the nanoscale would allow accurate analysis of their magnetic properties, which is crucial to understand and control new phases.

The negatively charged nitrogen-vacancy (NV) center in diamond exhibits a spin-1 triplet electronic ground state. The triplet state has energy levels sensitive to surrounding magnetic fields, and its spin states can be easily accessed via optical initialization/readout, and coherent microwave manipulation. This atomic-sized magnetometer is suitable to probe features of most  known 2D magnets with a dynamic range of magnetic field measurements spanning DC to several GHz and operational temperatures from below one to several hundreds of Kelvin\cite{Casola.2018}. Scanning magnetometry combining atomic force microscopy and NV center magnetometer allows for quantitative nanoscale imaging of magnetic fields and has been well established in room temperature measurements\cite{Balasubramanian.2008,Maletinsky.2012,Tetienne.2014,Haberle.2015,Chang.2017,Dovzhenko.2018}. While cyrogenic implementation of this technique is more challenging\cite{Pelliccione.2016,Thiel.2016}, it is crucial for the study of 2D magnets such as the chromium trihalides (CrX$_3$, X=Cl, Br, and I) as their magnetic ordering exists only at low temperature\cite{Kim.2019}. Recently, this technique has been implemented successfully to image  the magnetization in layered CrI$_3$ samples\cite{Thiel.2019}.

CrBr$_3$, for which ferromagnetic order persists from bulk crystal down to the monolayer, is a unique platform to study the ferromagnetism and spin fluctuation in the 2D limit\cite{Kim.2019b,Zhang.2019,jinchenhao.}. Although magnetic domains in layered CrBr$_3$ have been predicted from its anomalous hysteresis loop in magneto-photoluminescence and micromagnetometry measurements\cite{Zhang.2019,Kim.2019b}, the magnetic domain structure and its evolution has not been detected in real-space. Additionally, Zhang et al. detect magnetic domains only in multi-layer samples\cite{Zhang.2019}, while Kim et al. show signs of domains even in the monolayer case\cite{Kim.2019b}.This ambiguity cannot be resolved with microscale measurements.  In this work, we overcome this limitation using a single NV center in a diamond probe to map the stray magnetic field of the sample. We image the magnetic domain structures in a CrBr$_3$ bilayer, determine its magnetization, and study the magnetic domain evolution. In the main text, we focus on measurements of a bilayer CrBr$_3$ sample conducted with a diamond probe. Additional data obtained using a diamond probe with a different NV axis orientation, as well as the results from another CrBr$_3$ sample, are provided in the Supplementary Information.

\begin{figure}
\centering
\includegraphics{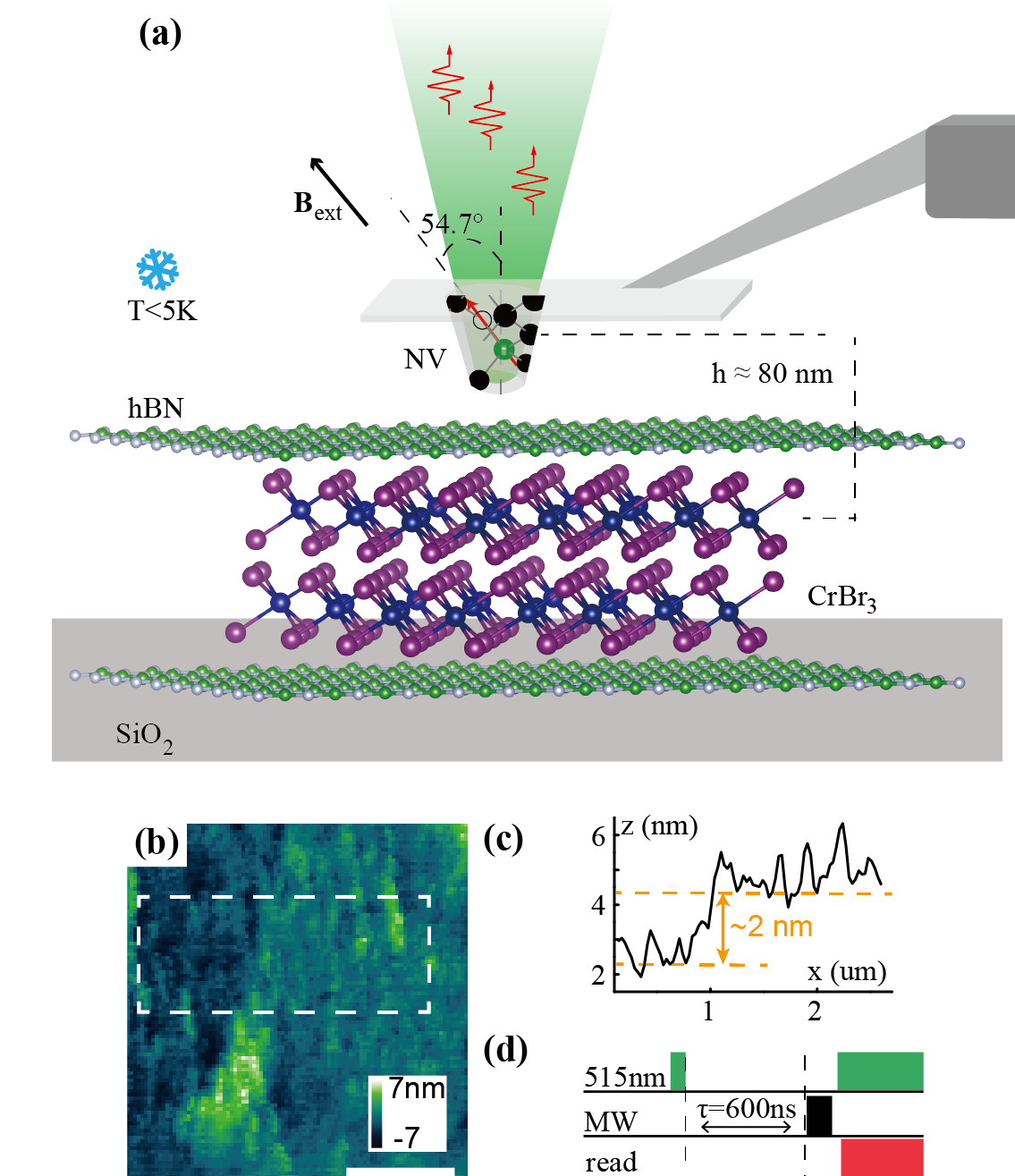}
\caption{\textbf{Cryogenic scanning magnetometry with a single NV center in the diamond tip.} \textbf{a},  Schematic of the experiment. The stray magnetic field of the CrBr$_3$ bilayer is measured using a single NV center in a diamond probe attached to the tuning fork of the AFM. The system is placed in a liquid Helium bath cryostat to maintain a temperature below 5~K during the measurement. \textbf{b} and \textbf{c}, A typical topography image of the measured area and the step at the edge of the sample, respectively. The scale bar is 1~$\mu$m. \textbf{d}, Measurement sequence of the pulsed optically detected magnetic resonance (ODMR).  }
\label{fig1}
\end{figure}

Fig.1~(a) shows a schematic of the scanning NV magnetometry setup. The NV center is implanted in the apex of the pillar etched from a diamond cantilever, which is attached to the tuning fork of an atomic force microscope (AFM). The  CrBr$_3$ sample, which is encapsulated with hexagonal boron nitride (hBN) on both sides, is transferred on a SiO$_2$/Si substrate in a glove-box filled with pure nitrogen (see Methods for details of sample preparation). The microscope head is suspended in an insertion tube filled with Helium buffer gas, which is dipped in the liquid Helium cryostat equipped with a set of vector superconducting coils. The NV spin state is optically measured through the cantilever supporting the diamond tip (see Methods). In the measurement, the AFM operates in a frequency modulation mode with a tip oscillation amplitude of about 1.5~nm. The topography of the sample is obtained from the AFM readout with a lateral spatial resolution of about 200~nm, which is limited by the diameter of the pillar apex.  A typical \textit{in-situ} AFM image of part of the CrBr$_3$ sample is shown in Fig.1~(b).  Fig.1~(c) shows the height of the step along the vertical direction by an average of the data in the dashed box in Fig.1~(b). The 2~nm step height indicates a bilayer sample.

The stray magnetic field is mapped by taking the electron spin resonance spectrum via the pulsed optically detected magnetic resonance (ODMR) scheme\cite{Dreau.2011} at each pixel. The measurement sequence is shown in Fig.1~(d). This pulsed scheme significantly decreases microwave heating compared with continuous wave ODMR\cite{Thiel.2019}. In our experiment, the sample temperature only increases by a few hundred milli-Kelvin from the base temperature of about 4.2~K during measurement. Moreover,  the magnetic field is measured via microwave pulses, which are applied 600~ns after the laser beam has been switched off. Thus, the measured stray magnetic field is not disturbed by laser-induced excitations such as spin-waves\cite{Cenker.2020,Zhang.2020}. Fig.2~(a) shows a typical stray magnetic field image of the sample under a 2~mT external magnetic field after being cooled down under zero field.  The external magnetic field is used to split the energy levels $\ket{m_s=\pm1}$ so that the direction of the stray magnetic field can be determined. The external magnetic field direction is set parallel to the NV axis in all measurements in this work to avoid the mixing of spin states due to an off-axis magnetic field component\cite{Tetienne.2012}. The axis of all the NV centers in the (100)-oriented diamond cantilevers we used here is about $54.7^\circ$ with respect to the vertical direction, as shown in Fig.1~(a). The pixel size in this work is set to 30~nm and the data accumulation time is 2~s at each pixel. The resulting stray magnetic field map clearly shows magnetic domains with prominent positive and negative values of the magnetic field and domain walls with nearly zero fields. To reveal further details, we reconstruct the magnetization from the stray magnetic field using reverse-propagation protocol\cite{Lima.2009,Dovzhenko.2018, Thiel.2019,Broadway.2020}.

\begin{figure}
\centering
\includegraphics{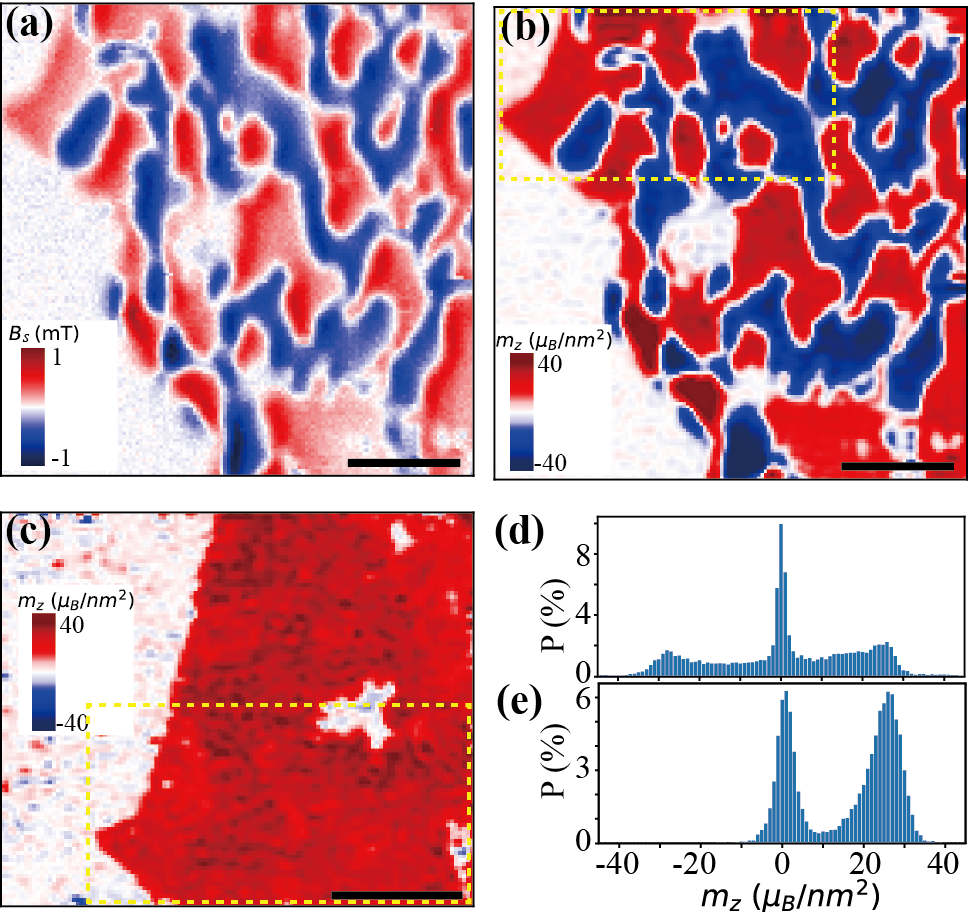}
\caption{\textbf{Magnetic domains and saturation magnetization.} 
\textbf{a}, Stray magnetic field and \textbf{b}, the reconstructed magnetization of a CrBr$_3$ bilayer under an external field of 2~mT along the NV axis. \textbf{c}, Magnetization image at external magnetic field of 11~mT. The dashed boxes in \textbf{b} and \textbf{c} denote the common sample area in the two images. Scale bar is 1~$\mu$m for all images. \textbf{d} and \textbf{e}, Histograms of the magnetization values in images \textbf{b} and \textbf{c}, respectively. }

\label{fig2}
\end{figure}

As discussed in more detail in the supplementary information, to uniquely determine the magnetization, we need some initial knowledge of the sample such as the direction of spin polarization.  It has been reported that few-layer CrBr$_3$ has an out-of-plane easy axis and can be polarized by a small external magnetic field of about 4~mT, while a relatively high external field ($B^c_\parallel\approx0.44$~T) is required to polarize the spins in the in-plane direction\cite{Kim.2019}. The in-plane external field component in this work is much lower than the critical field $B^c_\parallel$, allowing us to use the assumption of out-of-plane magnetization in the magnetization reconstruction. In addition, we neglect the finite thickness of the domain walls. Fig.2~(b) presents the magnetization image reconstructed from the stray magnetic field image in Fig.2~(a). It clearly shows the magnetic domain structure, with positive (negative) values indicating the magnetization direction parallel (anti-parallel) to the external magnetic field. The sample can be polarized by increasing the external magnetic field. Fig.2~(c) shows a magnetization image taken at 11~mT external magnetic field. The common areas in Figs.2~(b) and (c) are marked with dashed boxes. The saturation magnetization can be estimated using the magnetization statistics of the two magnetization images, as shown by the histograms in Figs.2~(d) and (e). Due to sample imperfection, measurement error, and truncation error in the reconstruction, the reconstructed magnetization is distributed in a range around the zero-magnetization and the saturation magnetization values. The near-zero-magnetization pixels are mostly in domain walls, defects, and the non-sample area on the left part of the images. The saturation magnetization values are $\sim 26 (-28)$ and $\sim 26 ~\mu_B$/nm$^2$, respectively, with $\mu_B$ the Bohr magneton. These values are close to the 3$\mu_B$  saturation moment per Cr$^{3+}$ ion in CrBr$_3$ at 0~K, i.e., $\sim 32~\mu_B$/nm$^2$ for a CrBr$_3$ bilayer\cite{Liu.2016}.

\begin{figure}
\centering
\includegraphics{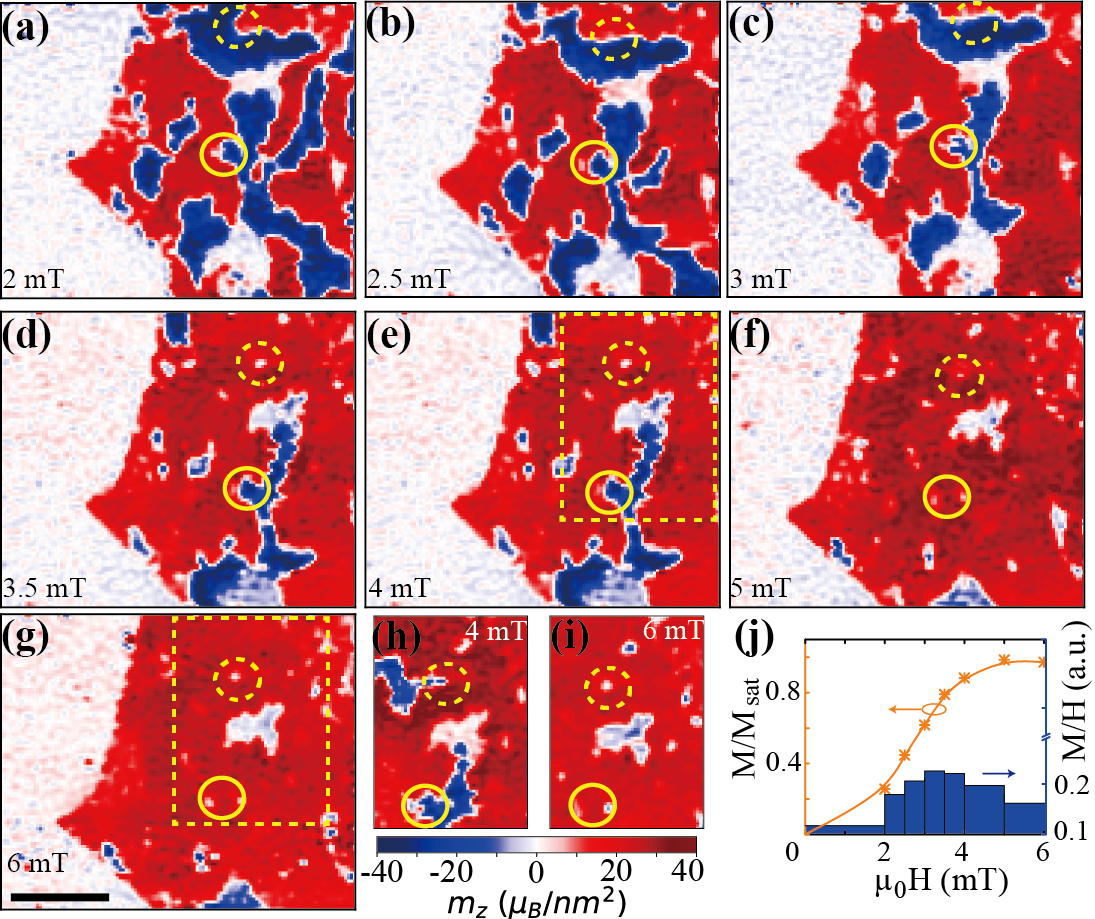}
\caption{\textbf{Magnetic domain evolution upon increasing the external magnetic field.} \textbf{a}-\textbf{g}, Magnetization images taken successively at external magnetic fields of 2, 2.5, 3, 3.5, 4, 5, and 6~mT along the NV axis, respectively. The sample is thermally demagnetized by heating to 45 K and then cooling down under zero field. \textbf{h} (\textbf{i}), Magnetization image of the sample area indicated by the dashed box in \textbf{e} (\textbf{g}) during another thermal cycle at external magnetic fields of 4 and 6~mT, respectively. The solid and dashed yellow circles denote the positions of two representative pinning sites. See the supplementary information for the other magnetization images.  Scale bar is 1~$\mu$m for all images. \textbf{j}, Initial magnetization curves extracted from the magnetization images in \textbf{a}-\textbf{g}. The blue bars are the ratios of magnetization to external magnetic field. }

\label{fig3}
\end{figure}

In addition to elucidating the magnetic domain structure of 2D magnets, our scanning magnetometry measurements enable a more detailed study of coercivity mechanisms in these systems. A multi-domain ferromagnet typically reverses its magnetization direction through processes such as nucleation of  reverse domains and their growth through domain wall motion\cite{Hubert.1998,Broadway.3182020}. Defects in the material alter the energy of the magnetic domain walls and hence affect domain wall motion. This behaviour can be demonstrated by taking magnetization images while varying the external magnetic field. Figs.3~(a)-(g) show magnetization images obtained while increasing the field from 2~mT to 6~mT after the sample is thermally demagnetized and cooled down under zero field. The area of positive (negative) domains grows (shrinks) with increasing field, as the domain walls move toward the negative domains. Before entirely disappearing, the negative domain size becomes very small, and only near-zero-magnetization spots of several tens of nanometer diameter are revealed in the magnetization images. As we are limited by the spatial resolution of the NV center (about 80 nm for the results shown in the main text, determined by the distance between the NV center and the sample for the diamond probe), we cannot obtain the detailed magnetization pattern inside the spots. These spots are usually associated with defects, which increase the local switching field. Domain walls are pinned by these defects (see Figs.3 (a)-(i), solid and dashed yellow circles). To confirm the pinning sites, we compare the magnetization images at 4 and 6 mT for different thermal cycles (see dashed box in Figs.3 (e) and (g), and compare to Figs.3 (h) and (i)). Though the magnetic domain structures are different upon successive thermal cycles, the positions of pinning sites are reproducible.

To verify that  the pinning effect is a dominant coercivity mechanism, we extract the initial magnetization curve of the thermally demagnetized sample by estimating its average magnetization as $M/M_{sat}=\frac{N_+-N_-}{N_++N_-}$, where $M_{sat}$ is the saturation magnetization and $N_+$ ($N_-$) is the number of pixels with evident positive (negative) magnetization (absolute value greater than 10 $\mu_B$/nm$^2$, according to the Fig.2 (e)). When pinning effects are negligible, the magnetic domain walls can move freely, resulting in a high initial magnetization even with a small external magnetic field. Defects, however,  increase the energy barrier for the displacement of  domain walls. In samples with a large defect density, the magnetization increases slowly until the external magnetic field is large enough to overcome the pinning energy. The initial magnetization curve shown in Fig.3~(j) is measured at external magnetic fields above 2~mT. We extend the curve to the origin using B-spline interpolation, assuming zero magnetization in a thermally demagnetized sample. The average permeability is very low under field of 2~mT and it significantly increases when the field is above 2~mT (see the blue bars in Fig.3 (j)), which is consistent with the behaviour of a pinning effect dominated initial magnetization.

\begin{figure}
\centering
\includegraphics{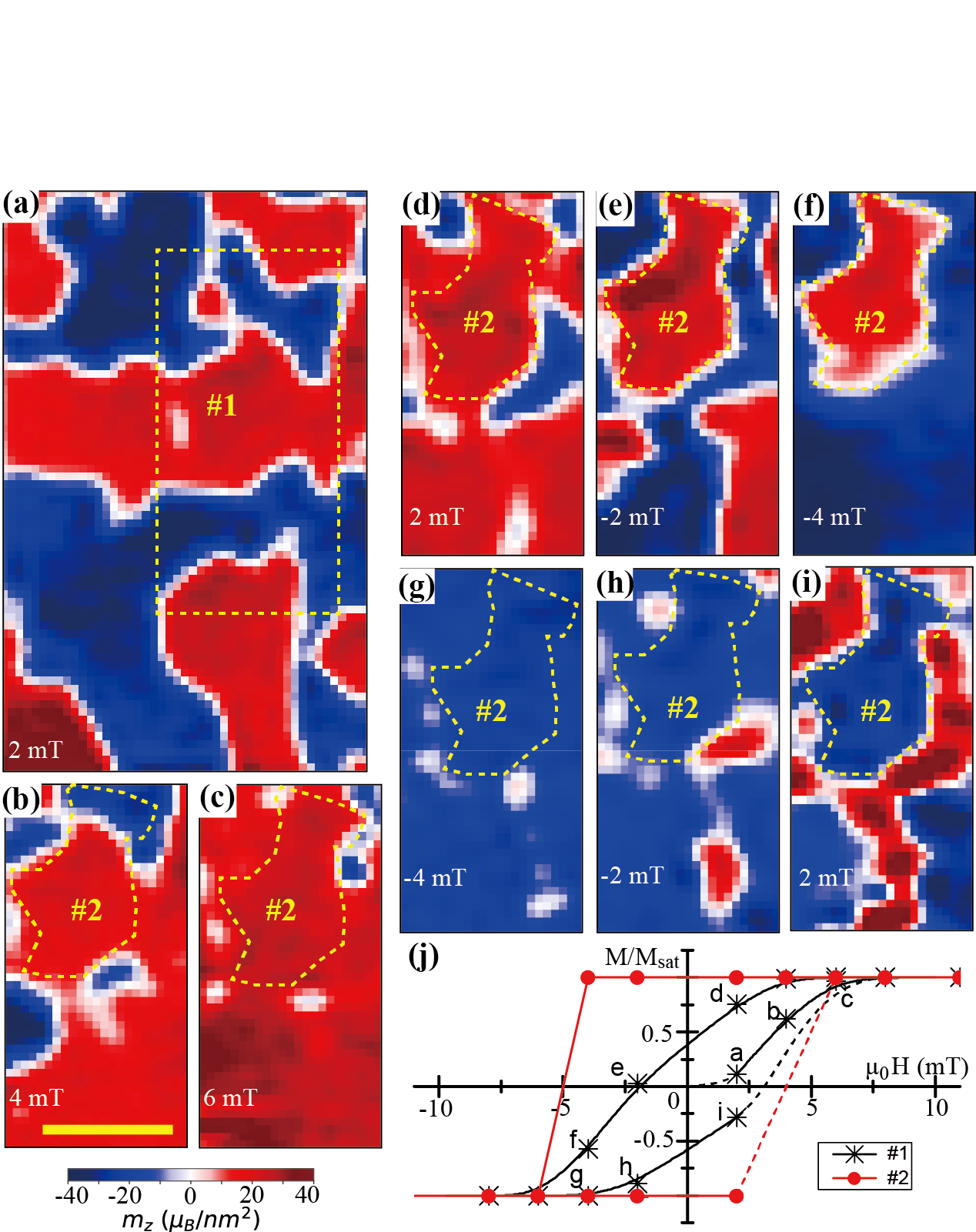}
\caption{\textbf{Magnetic hysteresis loop.} \textbf{a}, Magnetic domains at 2~mT external magnetic field along the NV axis after being thermally demagnetized and cooled down under zero field. The dashed box denotes the area $\#1$ used to analyze the hysteresis loop. \textbf{b}-\textbf{i}, Representative magnetization images of area $\#1$ on the hysteresis loop with the corresponding labels in \textbf{j}, and the dashed lines denote the area $\#2$. Scale bar is 1~$\mu$m for all the images. \textbf{j}, Hysteresis loop extracted from the magnetization images of area $\#1$ and $\#2$, which  are denoted by the black stars and red dots, respectively. The solid curve connects measured values, while the dashed curve is an extension to demagnetized and saturated states.
}

\label{fig4}
\end{figure}

We use a similar approach to measure the magnetic hysteresis loop.  Fig.4~(a) shows the magnetization image at an external magnetic field of 2~mT after thermal demagnetization. We choose the area marked by the dashed box ($\#1$) to analyse the magnetization as we cycle the external magnetic field to saturate the magnetization in both the positive and negative z direction. Representative magnetization images  are shown in Figs.4~(b)-(i) (see supplementary information for all images). The hysteresis loop of this area is shown by the black curve in Fig.4~(j).  Remarkably, a magnetic domain with the same irregular shape but opposite magnetization direction is observed in Figs.4~(e) and (i). The border of the magnetic domain ($\#2$) is denoted by the dashed line. We note that there are several defects (shown in Figs.4 (c) and (g)) around the $\#$2 region, leading to a very hard inverted domain. The magnetic domain wall is strongly pinned by the defects, and the magnetization is inverted  abruptly when the external field is larger than $\sim$5~mT. This can be interpreted as an increased local switching field due to a strong pinning effect, which is consistent with the higher coercive field indicated by the near-rectangular local hysteresis loop of this domain, as shown by the red curve in Fig.4~(j). The shape of the hysteresis loop is thus highly dependent on the area used to estimate the magnetization and on the local pinning sites. The rectangular hysteresis loop in bilayer CrBr$_3$ reported in a previous study\cite{Zhang.2019} could be attributed to this local pinning dependence.

In conclusion, we study  magnetic domains in few-layer samples of CrBr$_3$ by quantitatively mapping the stray magnetic field with a cryogenic scanning magnetometer based on a single NV center in a diamond probe. The magnetization of bilayer CrBr$_3$ is determined, and the magnetic domain evolution is observed in real space. We show that pinning is the dominant coercivity mechanism by observing the evolution of both the individual magnetic domains and the average magnetization with changing external magnetic field. To verify the observations, we reproduce similar magnetic-domain structures using micromagnetic simulation with parameters within the ranges estimated from measurement of CrBr$_3$ bulk\cite{Richter.2018} (see Methods and Supplementary Information). We find the hysteresis loop cannot be reproduced due to the limited capability of the micromagnetic simulation to account for the structure defects and pinning effect. This also support our conclusion that pinning effect is a dominant mechanism in the domain reversal of CrBr$_3$ bilayer. We note that our approach is also compatible with other pulsed measurement sequences which can be used to detect electron spin resonance\cite{Grinolds.2013}, nuclear magnetic resonance\cite{Haberle.2015} and spin waves\cite{vanderSar.2015} in the 2D magnetic materials.


\section*{Acknowledgements}
The authors thank  Dr. Thomas Oeckinghaus for his support with the experiment. J.W. acknowledges the Baden-W\"urttemberg Foundation, the European Research Council (ERC) (SMel grant agreement No. 742610). R.S. thanks the EU ASTERIQS. The work at U. Washington is mainly supported by DOE BES DE-SC0018171. Device fabrication is partially  supported by AFOSR MURI program, grant no. FA9550-19-1-0390. The authors also acknowledge the use of the facilities and instrumentation supported by NSF MRSEC DMR-1719797.

\clearpage

\textbf{Methods}

\textbf{Sample fabrication}
The hBN flakes of 10-30 nm were mechanically exfoliated onto 90 nm SiO2/Si substrates and examined by optical and atomic force microscopy under ambient conditions. Only atomically clean and smooth flakes were used for making samples. A V/Au (10/200 nm) microwave coplanar waveguide was deposited onto an 285 nm SiO$_2$/Si substrate using standard electron beam lithography with a bilayer resist (A4 495 and A4 950 polymethyl methacrylate (PMMA)) and electron beam evaporation. CrBr$_3$ crystals were exfoliated onto 90 nm SiO2/Si substrates in an inert gas glovebox with water and oxygen concentration less than 0.1 ppm. The CrBr$_3$ flake thickness was identified by optical contrast and atomic force microscopy. The layer assembly was performed in the glovebox using a polymer-based dry transfer technique. The flakes were picked up sequentially: top hBN, CrBr$_3$, bottom hBN. The resulting stacks were then transferred and released in a gap of the pre-patterned coplanar waveguide. In the resulting heterostructure, the CrBr$_3$ flake is fully encapsulated on both sides. Finally, the polymer was dissolved in chloroform for less than five minutes to minimize the exposure to ambient conditions. 

\textbf{Confocal microscope}
The optics of the confocal microscope consists of the low-temperature objective (Attocube LT-APO/VISIR/0.82) with 0.82 numerical aperture and home-built optics head (see supplementary information). The 515~nm excitation laser generated by an electrically driven laser diode is transmitted to the optics head through a polarization maintaining single-mode fiber and collimated by an objective lens. A pair of steering mirrors are used to align the beam for perpendicular incidence to the center of the objective. The NV center's fluorescence photons are also collected by the objective and transmitted through the same free-beam path to the optics head. The collected fluorescence photons are separated from the green laser beam via a dichroic mirror and then passed through a band-pass filter to further decrease background photons. Finally, the photons are coupled to a single-mode fiber and detected by a fiber-coupled single-photon detector.

\textbf{Stray magnetic field measurement}
With an external magnetic field applied along the NV-axis, $B_{\parallel}$, the spin states $\ket{m_s=0}$ and $\ket{m_s=\pm 1}$ exhibit Zeeman splitting of $f= D_s \pm \gamma_e B_{\parallel}$, where $D_s\approx2.87$~GHz is the zero field splitting between levels $\ket{m_s=\pm 1}$ and $\ket{m_s=0}$ and $\gamma_e=28$~GHz/T is the electronic spin gyromagnetic ratio.  In the pulsed ODMR measurement, the NV center is optically initialized in spin state $\ket{m_s=0}$. After a delay of $\tau=600$~ns, a $\pi$-pulse (about 80 ns) of microwave radiation is applied. If the microwave frequency is in resonance with one of the transitions, the NV center is driven to spin state $\ket{m_s=\pm 1}$. The population difference between the spin states $\ket{m_s=0}$ and $\ket{m_s=\pm 1}$ can then be optically read out via fluorescence contrast. In our experiment, the laser pulse duration is 600~ns, and only the first 400~ns of the fluorescence photon signal is used in the data analysis. The ODMR curve is obtained by sweeping the microwave frequency, which is achieved by modulating a microwave signal with a pair of sinusoidal signals in quadrature via an IQ mixer. All the control signals of the measurement sequences are generated using an arbitrary waveform generator (AWG) so that the measurement sequence is well synchronized, and fast microwave frequency sweeping is realized by altering the frequency of the sinusoidal signals in each unit segment.

\textbf{Micromagnetic simulation}
To complement the experimental results micromagnetic simulations of the systems magnetic ground state were conducted using ``MuMax3''\cite{Vansteenkiste.2014}. Saturation magnetization was set to $M_s$=270~kA/m, which is in accordance to the values measured here and reported in Ref[\cite{Richter.2018}]. Uniaxial magnetic anisotropy constant was assumed to be $K_u$=86 kJ/m$^3$ along the normal axis, which has been reported for bulk material\cite{Richter.2018}. A global exchange stiffness constant $A_{ex}$ was first roughly estimated from Curie temperature and experimentally observed domain wall width to lie within $10^{-12}$ to $10^{-14}$ J/m. Magnetization was initialized in a random configuration and then relaxed to the minimum energy state at zero external field. This was done for varying values of $A_{ex}$ from the interval estimated above. 

The results for the normal magnetization component are shown in the Supplementary Information for a system trying to locally approximate the irregular shape of the real sample. This simulation assumed $A_{ex} = 3\times10^{-13}$~J/m, which resulted in the closest match for the experiment. The simulation shows a domain structure that is qualitatively very similar to the experimental observations, which are further supported by this result. However, due to the limited capacity of the simulation to account for structural defects and pinning effects, the hysteresis behavior observed experimentally could not accurately be recreated by it. This further supports the conclusion that pinning is a major factor in this materials hysteresis.

\end{document}